\documentclass[12pt,centertags,a4paper]{amsart}
\pdfoutput=1
\usepackage{latexsym,eriktheorem,noans,times,harvard}
\usepackage{graphicx}%
\usepackage{bm,bbm}
\usepackage{amsmath,xcolor}
\usepackage{paralist} % for compactitem, etc.
\usepackage{multirow}
\swapnumbers \theoremstyle{dissprop} {\theoremstyle{dissdefi}
 }

{\theoremstyle{dissdefi} }

\newcommand{\bracket}[2]{\left\langle #1,#2 \right\rangle}

\DeclareMathOperator{\sgn}{sgn}
\DeclareMathOperator{\cov}{cov}
\begin{document}
\title[Quantifying Model Risk]{Quantifying the Model Risk Inherent in the Calibration and Recalibration of Option Pricing Models}
\author{Yu Feng}
\address[Yu Feng]{University of Technology Sydney, PO Box 123, Broadway, NSW 2007, Australia}
\email[Yu Feng]{Yu.Feng-5@student.uts.edu.au}
\author{Ralph Rudd}
\address[Ralph Rudd]{The African Institute for Financial Markets and Risk Management (AIFMRM), University of Cape Town}
\email[Ralph Rudd]{}
\author{Christopher Baker}
\address[Christopher Baker]{The African Institute for Financial Markets and Risk Management (AIFMRM), University of Cape Town}
\email[Christopher Baker]{}
\author{Qaphela Mashalaba}
\address[Qaphela Mashalaba]{The African Institute for Financial Markets and Risk Management (AIFMRM), University of Cape Town}
\email[Qaphela Mashalaba]{}
\author{Melusi Mavuso}
\address[Melusi Mavuso]{The African Institute for Financial Markets and Risk Management (AIFMRM), University of Cape Town}
\email[Melusi Mavuso]{}
\author{Erik Schl\"ogl}
\address[Erik Schl\"ogl]{University of Technology Sydney, PO Box 123, Broadway, NSW 2007, Australia}
\email[Corresponding author]{Erik.Schlogl@uts.edu.au}
\date{This version: \today. Part of the initial research for this paper was conducted at the Financial Mathematics Team Challenge 2016 at the University of Cape Town. We would like to thank Sam Cohen for helpful comments on earlier drafts of this paper. The usual disclaimers apply.}
\bibliographystyle{kluwer}
\begin{abstract}
We focus on two particular aspects of model risk: the inability of a chosen model to fit observed market prices at a given point in time (calibration error) and the model risk due to recalibration of model parameters (in contradiction to the model assumptions). In this context, we follow the approach of \citeasnoun{glasserman2014robust} and use relative entropy as a pre-metric in order to quantify these two sources of model risk in a common framework, and consider the trade--offs between them when choosing a model and the frequency with which to recalibrate to the market. We illustrate this approach applied to the models of \citeasnoun{OZ:Bla&Sch:73} and \citeasnoun{OZ:Heston:93}, using option data for Apple (AAPL) and Google (GOOG). We find that recalibrating a model more frequently simply shifts model risk from one type to another, without any substantial reduction of aggregate model risk. Furthermore, moving to a more complicated stochastic model is seen to be counterproductive if one requires a high degree of robustness, for example as quantified by a 99\% quantile of aggregate model risk.
\end{abstract} \maketitle
\markleft{YU FENG ET AL.}
\section{Introduction}
The renowned statistician George E. P. Box wrote that ``essentially, all models are wrong, but some are useful.''\footnote{See \citeasnoun{OZ:Box:87}.} This is certainly true in finance, where many models and techniques that have been extensively empirically invalidated remain in widespread use, not just in academia, but also (perhaps especially) among practitioners. At times, the way models are used directly contradicts the model assumptions: As observed market prices change, parameters in option pricing models, which are assumed to be time--invariant, are recalibrated, often on a daily basis. Incorrect models, and model misuse, represent a source of risk that is being increasingly recognised --- this is called ``model risk.'' As a paper by the Board of Governors of the Federal Reserve System put it in
2011,\footnote{See \citeasnoun{fedgov2011}.} ``The use of models invariably presents model risk, which is the potential for
adverse consequences from decisions based on incorrect or misused model outputs
and reports.''

In broad terms, one could identify four general classes of model risk inherent to the way mathematical models are used in finance, for example in (but not limited to) option pricing applications:
\begin{itemize}
\item Parameter uncertainty (and sensitivity to parameters) --- let's call this \textbf{``Type 0''} model risk for short. If model
parameters need to be statistically estimated, they will only be known up to
some level of statistical confidence, and this parameter uncertainty induces
uncertainty about the correctness of the model outputs.\footnote{Examples of where this type of risk is considered explicitly in the literature include \citeasnoun{CDO:Loeffler:03}, \citeasnoun{Ban&Sch:13} and \citeasnoun{Ker&Ber&Sch:10}.}
\item Inability to fit a model to a full set of simultaneous market observations
--- this is ``calibration error,'' let's call this \textbf{``Type 1''} model risk for short. To the extent that a model cannot match
observed prices on a given day, single-day (a.k.a. ``cross-sectional'') market
data already contradicts the model assumptions. The classical example of
this is the Black/Scholes implied volatility smile.
\item Change in parameters due to recalibration --- let's call this \textbf{``Type 2''} model risk for short. Once one moves
from one day to the next, this aspect of model risk becomes apparent:
In order to again fit the market as closely as possible, it is common practice in the industry to recalibrate models. This recalibration results in model parameters (which the models assume to be fixed) changing from day to day, contradicting the model assumptions.
\item The ``true'' dynamics of state variables don't match model dynamics\footnote{This type of model risk is considered for example in \citeasnoun{Ker&Ber&Sch:10}, who also relate this to \emph{identification risk}, which they define as risk which ``arises when observationally indistinguishable models have different consequences for capital reserves.''} --- let's call this violation
of model assumptions \textbf{``Type 3''} model risk.\footnote{\citeasnoun{Bou&Dan&Kou&Mai:14} present a method for making value--at--risk more robust with respect to this source of model risk by ``learning'' from the results of model backtesting.} The classical example of this is the econometric rejection of the hypothesis that asset prices follow geometric Brownian motion, thus invalidating the key assumption in the seminal model of \citeasnoun{OZ:Bla&Sch:73}. This type of model risk would impact in particular the effectiveness of hedging strategies based on a model.\footnote{\citeasnoun{Det&Pac:16} take the approach of measuring model risk based on the residual profit/loss from hedging in a misspecified model.}
\end{itemize}
Note that there is a gradual transition between the different types of model risk, and depending on one's modelling choices, to a certain extent one can trade off one type of model risk against another. For example,
\begin{itemize}
\item  Less stringent requirements of an exact fit to market observations (Type 1)
allows less frequent recalibration (Type 2).
\item Instead of different model dynamics (Type 3), one could consider a parameterised
family of models (Type 2).
\item Regime--switching models ``legalise'' changes in parameters, so Type 2 becomes
more like Type 3.
\item Adding parameters shifts model risk from Type 1 to Type 2 (or, to a certain extent, to Type 0).
\item Adding state variables shifts model risk from Type 2 to Type 3.
\end{itemize}
\citeasnoun{glasserman2014robust} propose relative entropy as a consistent pre-metric by which to measure model risk from different sources.\footnote{Instead of using a relative entropy pre-metric, one could approach quantifying model risk in terms of optimal--transport distance, using for example Wasserstein distance, which most recently has become popular for this purpose (see \citeasnoun{Bar&Dra&Tan:18}, \citeasnoun{Bla&Che&Zho:18} and \citeasnoun{Fen&Sch:18}). In the present paper, we follow the more established approach using relative entropy, which has its roots in the seminal work of Hansen and Sargent (see e.g.  \citeasnoun{Han&Sar:06}).}
What matters in the application of mathematical models in finance is the probability distributions which the models imply,\footnote{\citeasnoun{Bre&Csi:16} call this \emph{distribution model risk}.} either under a ``risk--neutral'' probability measure (for applications to relative pricing of financial instruments) or the ``physical'' (a.k.a. ``real--world'') probability measure (for risk management applications such as the calculation of expected shortfall). Each type of model risk manifests itself as some form of ambiguity about the ``true'' probability measures which should be used for these purposes, and being able to quantify different types of model risk in a unified setting using a pre-metric for the divergence between distributions (like relative entropy) allows one to make an informed choice about the trade--offs between different sources of model risk. \citeasnoun{glasserman2014robust} postulate a ``relative entropy budget'' defining a set of models sufficiently close (in the sense of relative entropy) to a nominal reference model to be considered in an evaluation of model risk expressed as a ``worst case'' expectation --- i.e., a worst--case price or a worst--case risk measure. However, they say little as to how one typically would obtain a specific number for this ``relative entropy budget''. In a sense, we invert this problem by noting that higher relative entropy between model distributions indicates higher model risk, and propose a method to jointly evaluate model risk of two types, based on how this model risk manifests itself when option pricing models are calibrated and recalibrated to liquid market instruments.

We focus on the model risk inherent in the calibration and recalibration (i.e., in the above terminology, Types 1 and 2) of option pricing models, and to illustrate our approach we consider the models of \citeasnoun{OZ:Bla&Sch:73} and \citeasnoun{OZ:Heston:93}, thus comparing the most classical option pricing model with its popular extension incorporating stochastic volatility. Clearly, if (as is often the case in practice) one focuses solely on calibration error, \citeasnoun{OZ:Heston:93} will always be preferred to \citeasnoun{OZ:Bla&Sch:73}, and more frequent recalibration preferred to less. We quantify calibration and recalibration risk in both models applied to equity option data, and also explore the trade--off between these two types of model risk, finding that there is no longer a trivial answer to the question which model and which recalibration frequency should be preferred when these two sources of model risk are considered in a unified framework.

The rest of the paper is organised as follows. Section \ref{framework} introduces a framework for the joint evaluation of model risk due to calibration error and due to model recalibration. The numerical implementation of the method is discussed in Section \ref{numerical}. Section \ref{results} presents the results obtained by applying this method to option price data, and Section \ref{conclusion} concludes.

\section{Calibration error, model risk due to recalibration, and treatment of latent state variables}\label{framework}
As noted above, model risk is reflected in the ambiguity with regard to the ``correct'' probability distribution to use for relative pricing or risk assessment. Following \citeasnoun{glasserman2014robust}, we quantify this ambiguity using the divergence between probability measures. In the present context, these can be classified as divergence measures defined as a function $D(\cdot||\cdot):~S\times S\to\mathbb{R}$ satisfying
\begin{align}\label{eq:divergence}
D(Q||P)\geq0&~~ ~~\forall P, Q\in S\\
D(Q||P)=0&~\Leftrightarrow P=Q
\end{align}
where $S$ is a space of all probability measures with a common support. More specifically, most divergence measures belong to the class of $f$-divergence, which gives the divergence between two equivalent measures as:\footnote{See e.g. \citeasnoun{ali1966general}, \citeasnoun{csisz1967information} or \citeasnoun{ahmadi2012entropic}.}
\begin{align}\label{eq:gentropic}
D(Q||P)=\left.\int_\Omega f\left(\frac{dQ}{dP}\right)dP\right.
\end{align}
where $f$ is a convex function of the Radon-Nikodym derivative satisfying $f(1)=0$.
Kullback--Leibler divergence (a.k.a. relative entropy) is the most common $f$-divergence, which assigns $f(m)=m\ln m$. It is noted that the methodology of this paper applies to all types of statistical distances in principle, though in the empirical study the Kullback--Leibler divergence is adopted due to its simplicity and widespread use.

If we wish to quantify calibration error (Type 1 model risk) in this fashion, then in equations (\ref{eq:divergence})--(\ref{eq:gentropic}), the probability measure $P$ corresponds to the calibrated model and thus is parametric in some form. The probability measure $Q$, on the other hand, serves as a reference measure exactly matching observed market prices at a given point in time, unrestricted by the assumptions of the model under consideration. On calibrating an option pricing model, we may regard the measure $Q$ as some non-parametric risk--neutral measure that explains the market in full assuming absence of arbitrage.
In practice, however, the measure $Q$ is not unique as the market is usually incomplete.
We therefore define the space of all probability measures that explains the market in full by $S_{Q}\subset S$.

We may further define the space of probability measures given by all possible choices of parameter values for the target model by $S_{P}\subset S$.
The new calibration methodology proposed here aims to minimise the calibration error as quantified by the divergence between the two measures $P$ and $Q$, taken from their respective spaces, i.e.
\begin{align}\label{eq:pair}
(Q^*,P^*)=\arg\min_{Q\in S_{Q}, P\in S_{P}}D(Q||P)
\end{align}
This is to say, the new approach attempts to calibrate a model measure $P^*$ (i.e., a set of model parameters $\theta^*$) and non-parametric perfect fit to the market (at a given point in time) $Q^*$, in a fashion which minimises the calibration error expressed by
\begin{align}
\eta_1=D(Q^*||P^*)
\end{align}
This is not an end in itself --- it is required in order to compare model risk due to calibration error and model risk due to recalibration (as specified below) in a unified framework.

The classical approaches of model calibration, such as minimising the mean--squared error between model and market prices for options, would be inappropriate in this context, as they would lead to unnecessarily high model risk quantities. It is the choice of divergence measure which informs the calibration procedure, resulting in a pair of probability measures, $(Q^*,P^*)$, one of which corresponds to the calibrated model while the other provides a consistent reference measure fitting the market exactly.

To quantify the model risk due to recalibration, let us consider the more specific case where the model is Markovian in a vector of observable state variables $X$, the model is characterised by a vector of model parameters $\theta$, and market prices are given for European option prices of a single maturity $T$.\footnote{This last assumption of a single maturity $T$ avoids the need to constrain the choice of $Q$ to ensure the absence of calendar spread arbitrage between non-parametric risk--neutral measures for different time horizons --- parametric models typically ensure this by construction. If we appropriately constrain $Q$, this assumption can be lifted.} Suppose we solved (\ref{eq:pair}) yesterday (at time $t_{i-1}$) to obtain a $P^*$ --- to be as explicit as possible, denote this by
\begin{equation}\label{eq:Pnotation}
P_{t_{i-1},X_{t_{i-1}}(\omega),\theta^*_{t_{i-1}}}(A)\qquad\forall\ A\in\mathcal{F}_T
\end{equation}
I.e., this is a (conditional) probability measure defined on all $\mathcal{F}_T$--measurable events, where the conditioning is on the state variables at time $t_{i-1}$, $X_{t_{i-1}}$, and we write $X_{t_{i-1}}(\omega)$ to express that the time $t_{i-1}$ realisations of the state variables are known at the time that these probabilities are evaluated. We write the subscript $\theta^*_{t_{i-1}}$ to express that these probabilities are evaluated in a model with parameters calibrated by solving (\ref{eq:pair}) at time $t_{i-1}$. Furthermore, denote the non-parametric measure $Q^*$ resulting from solving (\ref{eq:pair}) at time $t_{i-1}$ by $Q_{t_{i-1}}$.

Now, if we recalibrate today (at time $t_i$) by solving (\ref{eq:pair}), we obtain $Q_{t_i}$ and
\begin{equation}
P_{t_{i},X_{t_{i}}(\omega),\theta^*_{t_{i}}}(A)\qquad\forall\ A\in\mathcal{F}_T
\end{equation}
We can then define the model risk quantity due to recalibration as
\begin{equation}\label{eq:eta2}
\eta_2=D(P_{t_{i},X_{t_{i}}(\omega),\theta^*_{t_{i}}}||P_{t_{i},X_{t_{i}}(\omega),\theta^*_{t_{i-1}}})
\end{equation}
which is the divergence between the (conditional) probability measures evaluated at time $t_i$, where one measure is based on the recalibrated parameters $\theta^*_{t_{i}}$ and the other is based on the previously calibrated parameters $\theta^*_{t_{i-1}}$ (thus expressing, in terms of divergence, the inconsistency with the model assumptions due to the fact that we are going ``outside of the model'' to change parameters in recalibration). The aggregate of calibration error and model risk due to recalibration is then
\begin{equation}
\eta_3=D(Q_{t_{i}}||P_{t_{i},X_{t_{i}}(\omega),\theta^*_{t_{i-1}}})
\end{equation}
i.e., the divergence between the non-parametric probability measure $Q_{t_i}$ obtained by solving (\ref{eq:pair}) at time $t_i$, and the non-recalibrated parametric probability measure, consisting of probabilities conditional on the state at time $t_i$, but based on model parameters obtained by solving (\ref{eq:pair}) at time $t_{i-1}$. However, this approach minimises the divergence between the reference distribution and the recalibrated distribution, thus arguably overstating the divergence to the non-recalibrated (i.e. model--consistent) distribution, and therefore overstating the aggregate model risk $\eta_3$.

Alternatively, we may choose as the non-parametric reference distribution at time $t_i$:
\begin{equation}\label{eq:qhat}
\hat Q_{t_i}=\text{arg}\min_{Q\in S_Q}D(Q||P_{t_{i},X_{t_{i}}(\omega),\theta^*_{t_{i-1}}})
\end{equation}
resulting in a lower aggregate model risk of
\begin{equation}
\eta_3=D(\hat Q_{t_{i}}||P_{t_{i},X_{t_{i}}(\omega),\theta^*_{t_{i-1}}})
\end{equation}
Note that $\theta^*_{t_i}$ is still obtained by solving (\ref{eq:pair}), because both $Q_{t_i}$ and $\hat Q_{t_i}$ represent non-parametric probability measures fitting observed market prices exactly, so $\theta^*_{t_i}$ remains the best available parametric fit to the market at time $t_i$ ($\hat Q_{t_i}$ is only used to determine minimum divergence of the non-recalibrated model to a measure giving a perfect fit).

In the heuristic schematic of Figure 1(a),\footnote{Note that these graphs are for the purpose of heuristic illustration only --- in particular, we are not requiring that the two sets of probability measures are convex.} point A represents $P_{t_{i},X_{t_{i}}(\omega),\theta^*_{t_{i}}}$, being the parametric probability measure ``closest'' to the set of non-parametric probability measures fitting the market exactly, where point C represents $Q_{t_i}$. If we do not recalibrate at time $t_i$, we end up with the parametric probability measure $P_{t_{i},X_{t_{i}}(\omega),\theta^*_{t_{i-1}}}$ (point B), to which $\hat Q_{t_i}$ (point D) is the ``closest'' non-parametric probability measure fitting the market exactly.

In the case of Kullback--Leibler divergence, note that if Type 1 (calibration error) and 2 (recalibration) model risk involve independent Radon--Nikodym derivatives, then, in the first case considered above, aggregate model risk equals the sum of the two components. In fact, the Radon-Nikodym derivatives, as random variables, take the key role in evaluating the two types of model risk. At the time the model is recalibrated, we again consider the optimisation (\ref{eq:pair}), with $S_Q$ now changed to reflect the change in observed market prices, so we have the following Radon--Nikodym derivatives:
\begin{align}
\text{For calibration error: }\qquad m_1=&\frac{dQ_{t_i}}{dP_{t_{i},X_{t_{i}}(\omega),\theta^*_{t_{i}}}}\label{eq:calerr}\\
\text{For model risk due to recalibration: }\qquad m_2=&\frac{dP_{t_{i},X_{t_{i}}(\omega),\theta^*_{t_{i}}}}{dP_{t_{i},X_{t_{i}}(\omega),\theta^*_{t_{i-1}}}}\\
\text{For aggregate model risk: }\qquad m_1m_2=&\frac{dQ_{t_i}}{dP_{t_{i},X_{t_{i}}(\omega),\theta^*_{t_{i-1}}}}\label{eq:aggerr}
\end{align}
Abbreviating $dP_{t_{i},X_{t_{i}}(\omega),\theta^*_{t_{i-1}}}$ as $P$ and $dP_{t_{i},X_{t_{i}}(\omega),\theta^*_{t_{i}}}$ as $P^*$, the aggregate risk can be expressed in terms of $m_1$ and $m_2$ as:
\begin{align} \label{eq:add}
	\eta_3%=&E[\frac{dQ^*}{dP^*}
	=&E^P[m_1m_2\ln(m_1m_2)]\\
	=&E^{P^*}[m_1\ln(m_1)]+E[m_1m_2\ln(m_2)]\\
	=&\eta_1+%E^{f}(m_2)E^{f}[m_1\ln(m_1)]+cov^f[m_2,m_1\ln(m_1)]\\
	[1-\cov^P(m_1,m_2)]\eta_2+\cov^P[m_1,m_2\ln(m_2)]
\end{align}
If $m_1$ and $m_2$ are independent, $\cov^P(m_2,m_1)=\cov^P[m_2,m_1\ln(m_1)]=0$. The total model risk is equal to the sum of the calibration risk and the recalibration risk. Surprisingly, in our empirical exploration below we found that this equality is followed quite well by the Black-Scholes model. However, it typically does not hold in the Heston model, suggesting substantial dependence (of Radon-Nikodym derivatives) between the calibration error and model risk due to recalibration.

\begin{figure}[t]
\begin{center}
\includegraphics[width=.45\textwidth]{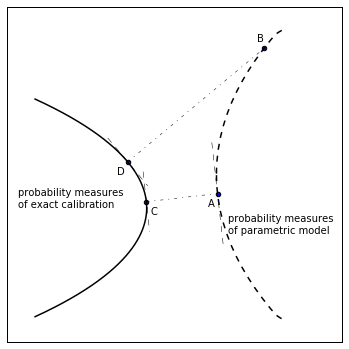}
\includegraphics[width=.45\textwidth]{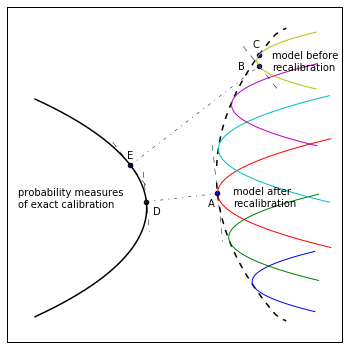}
\caption{Graphic illustration of the mathematical definitions of model risks for (a) observable state variable, (b) latent state variable.}\label{fig:2}
\end{center}
\end{figure}
We also consider models which involve one or more latent state variables. An example of that is the class of stochastic volatility models where the volatility is taken as a latent state variable rather than a model parameter (in the empirical examples below, we specifically consider the model of \citeasnoun{OZ:Heston:93}, which falls into this category). Under the framework of a single stochastic volatility state variable, a model specified by a given set of parameters forms a one-dimensional manifold (Fig.~\ref{fig:2}(b)) for possible realisations of the state variable, rather than a point in the Black-Scholes world (Fig.~\ref{fig:2}(a)).

Thus, the model which we are now considering is Markovian in a vector of state variables $(X,V)$, where the state variables $X$ are observable and the state variables $V$ are latent (unobservable). Then, the initial calibration problem (\ref{eq:pair}) becomes
\begin{align}\label{eq:min2}
(Q^*,v^*,\theta^*)=&\arg\min_{Q\in S_Q, v\in S_v, \theta\in S_\theta}D(Q||P(v;\theta))
\end{align}
where $S_v$ and $S_\theta$ are the sets of legitimate values of the state variables and the parameters, respectively. $\theta^*$ is the set of model parameters calibrated to the market, and $v^*$ is the best estimate of the latent state variables under the calibrated model.\footnote{This effectively treats the latent (unobserved) state variable as an additional parameter to be calibrated, but the recalibration of which does not contribute to (Type 2) model risk due to recalibration, because it is consistent with the model assumptions for this latent state variable to evolve stochastically. This does shift Type 2 model risk to Type 3, the risk that the state variable dynamics are not (econometrically) consistent with the dynamics assumed in the model. However, in the present paper we deliberately set aside Type 3 model risk for the purposes of our analysis, leaving the integration of all four types of model risk for future research.} The notation in (\ref{eq:Pnotation}) is amended to
\begin{equation}\label{eq:Pvnotation}
P_{t_{i-1},X_{t_{i-1}}(\omega),v^*_{t_{i-1}},\theta^*_{t_{i-1}}}(A)\qquad\forall\ A\in\mathcal{F}_T
\end{equation}
At time $t_i$, we have for the calibration error
\begin{equation}
\eta_1=D(Q_{t_i}||P_{t_{i},X_{t_{i}}(\omega),v^*_{t_{i}},\theta^*_{t_{i}}})
\end{equation}
The model risk due to recalibration is
\begin{equation}
\eta_2=D(P_{t_{i},X_{t_{i}}(\omega),v^*_{t_{i}},\theta^*_{t_{i}}}||P_{t_{i},X_{t_{i}}(\omega),v^*_{t_{i}},\theta^*_{t_{i-1}}})
\end{equation}
The aggregate model risk using $Q_{t_i}$ from $Q^*$ and $v^*_{t_{i}}$ from $v^*$ in (\ref{eq:min2}) is
\begin{equation}
\eta_3=D(Q_{t_i}||P_{t_{i},X_{t_{i}}(\omega),v^*_{t_{i}},\theta^*_{t_{i-1}}})
\end{equation}
or alternatively, using $\hat Q_{t_i}$ and $\hat v_{t_i}$ determined analogously to (\ref{eq:qhat}), i.e.,
\begin{align}
(\hat Q_{t_i},\hat v_{t_i})=&\arg\min_{Q\in S_Q, v\in S_v}D(Q||P_{t_i,X_{t_i}(\omega),v,\theta^*_{t_{i-1}}})
\end{align}
which results in
\begin{equation}
\hat\eta_3=D(\hat Q_{t_i}||P_{t_{i},X_{t_{i}}(\omega),\hat v_{t_{i}},\theta^*_{t_{i-1}}})
\end{equation}
We then have the following Radon--Nikodym derivatives:
\begin{align}
\text{For calibration error: }\qquad m_1=&\frac{dQ_{t_i}}{dP_{t_{i},X_{t_{i}}(\omega),v^*_{t_{i}},\theta^*_{t_{i}}}}\label{eq:vcalerr}\\
\text{For model risk due to recalibration: }\qquad m_2=&\frac{dP_{t_{i},X_{t_{i}}(\omega),v^*_{t_{i}},\theta^*_{t_{i}}}}{dP_{t_{i},X_{t_{i}}(\omega),v^*_{t_{i}},\theta^*_{t_{i-1}}}}\\
\text{For aggregate model risk: }\qquad m_1m_2=&\frac{dQ_{t_i}}{dP_{t_{i},X_{t_{i}}(\omega),v^*_{t_{i}},\theta^*_{t_{i-1}}}}\label{eq:vaggerr}
\end{align}
Note that the key difference between (\ref{eq:calerr})--(\ref{eq:aggerr}) and (\ref{eq:vcalerr})--(\ref{eq:vaggerr}) is that the change in $v$, being permitted by the model assumptions, does not contribute to the model risk quantities. In (\ref{eq:pair}) and (\ref{eq:min2}), we are deliberately prioritising the minimisation of calibration error, as this is congruent to the (often exclusive) focus of practitioners on calibration error (with little or no regard to model risk due to recalibration). If desired, one could reformulate this approach to prioritise the minimisation of aggregate model risk, or of model risk due to recalibration.

\section{Numerical implementation}\label{numerical}
In this section, we outline the numerical scheme for solving the minimisation problems arising when taking into account calibration error and model risk due to recalibration in the manner described in the previous section, including problems of the type (\ref{eq:pair}) involving the optimal choice of two probability measures. In this case, an iterative process is required, optimising two probability measures $Q$ and $P$ in turn until convergence, in the following manner:
\begin{description}
\item[1)] Produce $P^{(0)}$ from a parametric model based on an initial guess of the model parameters (and latent state variables, where required).
\item[2)] Solve for $Q^{(0)}$ via Lagrange multipliers for the constrained problem that minimises $D(Q^{(0)}||P^{(0)})$.
\item[3)] Solve for $P^{(1)}$ to obtain model parameters for the $P^{(1)}$ that minimises $D(Q^{(0)}||P^{(1)})$.
\item[4)] Iterate steps 2 and 3: $P^{(n)}\to Q^{(n)}\to P^{(n+1)}$ until convergence.
\end{description}
In Step 1, the initial guess may be obtained in several different ways. A common way is to minimise the mean--squared error between model and market option prices at all available strikes. We opted for the Broyden/Fletcher/Goldfarb/Shanno (BFGS) algorithm for conducting this initial calibration of the model parameters and (where required) latent state variables.

In Step 2, we solve the following constrained minimisation problem using Lagrange multipliers:
\begin{align}
Q^{(0)}&=\arg\min_{Q\in S_Q}D(Q||P^{(0)})\label{eq:prob}\\
s.t. &~\bm{B}\leq E^Q(\bm{Z})\leq\bm{A}\label{eq:cons}
\end{align}
Note that here we specify the constraints in the form of expectations under the measure $Q$, where these expectation are the model prices for our calibration instruments for the model based on the non-parametric reference distribution $Q$. In general, $\bm{B}$, $\bm{Z}$ and $\bm{A}$ are vectors; thus (\ref{eq:cons}) is a ``stack'' of inequality constraints representing observed market prices. Also notice that for generality we ``relax'' each equality constraint into two inequality constraints.  This is in order to account for the bid-ask spread of each option traded on the market. The vector $\bm{B}$ denotes a list of bid prices %for all the options traded on the market,
while the vector $\bm{A}$ contains ask prices. In a simplified scenario, where exact option prices are given, we may set $\bm{B}=\bm{A}$. $\bm{Z}$ denotes the vector of discounted option payoffs.
By introducing vectors of Lagrange multipliers $\boldsymbol{\lambda}_B$ and $\boldsymbol{\lambda}_A$, we convert the constrained problem to an unconstrained dual problem,
\begin{align}\label{eq:dual}
&\max_{\boldsymbol\lambda_B\geq 0, \boldsymbol\lambda_A\leq0}\min_{Q\in S_Q}D(Q||P^{(0)})-\boldsymbol\lambda_B\left[E^Q(\bm{Z})-\bm{B}\right]-\boldsymbol\lambda_A\left[E^Q(\bm{Z})-\bm{A}\right]
\end{align}

In the case of Kullback-Leibler divergence, solving the inner problem gives the probability density function $q^{(0)}$ of $Q^{(0)}$ in terms of the density $p^{(0)}$ of $P^{(0)}$,
\begin{align}\label{eq:pmelambda}
q^{(0)}=\frac{p^{(0)}e^{\boldsymbol\lambda \mathbf Z}}{E^{P^{(0)}}\left(e^{\boldsymbol\lambda \mathbf Z}\right)}
\end{align}
Substituting (\ref{eq:pmelambda}) into (\ref{eq:dual}), we get a maximisation problem,
\begin{align}
\max_{\boldsymbol\lambda_B\geq 0, \boldsymbol\lambda_A\leq0}
-%\frac{E^{P^{(0)}}\left(\mathbf Ze^{\boldsymbol\lambda \mathbf Z }\right)}
\ln {E^{P^{(0)}}\left(e^{(\boldsymbol\lambda_B+\boldsymbol\lambda_A) \mathbf Z}\right)}+(\boldsymbol\lambda_B+\boldsymbol\lambda_A)\frac{\boldsymbol B+\boldsymbol A}{2}- (\boldsymbol\lambda_{B}-\boldsymbol\lambda_A)\frac{\boldsymbol A-\boldsymbol B}{2}
\end{align}
If $\boldsymbol B=\boldsymbol A$, then the last term vanishes, representing the problem with exact market prices. If (component-wise) $\boldsymbol B>\boldsymbol A$, then the last term reflects a penality on the objective function that is proportional to the difference of the two Lagrange multipliers. We may therefore transform the Lagrange multipliers by
\begin{align}
\boldsymbol\lambda_+=\boldsymbol\lambda_B+\boldsymbol\lambda_A\\
\boldsymbol\lambda_-=\boldsymbol\lambda_B-\boldsymbol\lambda_A
\end{align}
and the objective function becomes
\begin{align}
\max_{\boldsymbol\lambda_+\in\mathbb{R}^d, \boldsymbol\lambda_-\geq|\boldsymbol\lambda_+|}
&-\ln {E^{P^{(0)}}\left(e^{\boldsymbol\lambda_+ \mathbf Z}\right)}+\boldsymbol\lambda_+\frac{\boldsymbol B+\boldsymbol A}{2}- \boldsymbol\lambda_-\frac{\boldsymbol A-\boldsymbol B}{2}\\=
\max_{\boldsymbol\lambda_+\in\mathbb{R}^d}
&-\ln {E^{P^{(0)}}\left(e^{\boldsymbol\lambda_+ \mathbf Z}\right)}+\boldsymbol\lambda_+\frac{\boldsymbol B+\boldsymbol A}{2}- |\boldsymbol\lambda_+|\frac{\boldsymbol A-\boldsymbol B}{2}
\end{align}
We may numerically solve the maximization problem by taking its gradient with respect to $\boldsymbol\lambda_+$,
\begin{align}\label{eq:newton}
E^{P^{(0)}}\left(\mathbf Ze^{\boldsymbol\lambda_+ \mathbf Z }\right)=\left[\frac{\boldsymbol B+\boldsymbol A}{2}-\bm{sgn}(\boldsymbol\lambda_+)\frac{\boldsymbol A-\boldsymbol B}{2}\right] E^{P^{(0)}}\left(e^{\boldsymbol\lambda_+ \mathbf Z}\right)
\end{align}
where the element-wise sign function $\bm{sgn}(\boldsymbol\lambda_+)$ assigns 1, -1 or 0 to each element of $\boldsymbol\lambda_+$.
However, due to the discontinuouity of the sign function (\ref{eq:newton}) cannot be solved directly in a stable way. To bypass this problem, we approximate the sign function with a continuous step function:
\begin{align}
\sgn(x)=1-\frac{2}{1+\exp\left(\frac{x}{\delta}\right)}
\end{align}
We use Powell's hybrid method to solve the multidimensional equations (\ref{eq:newton}), where $\delta$ controls the steepness of the function and carefully choosing this value is critical for a fast and stable convergence of the method.

In Step 3, we use L-BFGS-B algorithm to minimise the divergence with respect to model parameters (or latent variables or both). Step 2 and Step 3 are repeated until convergence. The convergence criterion adopted here is that all the percentage changes of parameters after one iteration do not exceed a certain threshold, say 0.1\%.

\section{Examining the trade--off between calibration error and model risk due to recalibration}\label{results}
As an application example of the method described in the previous two sections, we consider historical data consisting of daily market prices for call options on AAPL and GOOG stock over a period from 6 January 2004 to 19 December 2008 for AAPL and 4 January 2005 to 19 December 2008 for GOOG. This gives us a reasonably straightforward application example free of extraneous complications,\footnote{Although these options are of the American type, i.e. permitting early exercise, AAPL and GOOG did not pay any dividends during this period. Thus the possibility of early exercise may be ignored (see \citeasnoun{OZ:Merton:73}).} while still covering reasonably liquid options and including a period of ``interesting'' market volatility (2007/8). From this data, we remove options very far away from the money, restricting the range of strikes from delta 2.5\% to delta 97.5\%. Furthermore, we remove prices of options which had zero trading volume on a given day, in order to avoid using prices which are likely to be stale.

On this data we consider two parametric models, \citeasnoun{OZ:Bla&Sch:73} and \citeasnoun{OZ:Heston:93} --- arguably the two most popular option pricing models available, where the latter introduces a latent variable for stochastic volatility. The unified methodology, quantifying calibration error, model risk due to recalibration, and the aggregate of the two, allows us to explore the trade--off between calibration error (which is, unsurprisingly, reduced by moving from \citeasnoun{OZ:Bla&Sch:73} to \citeasnoun{OZ:Heston:93}) and model risk due to recalibration (which has hitherto been largely ignored) when moving from one parametric model to another as well as when changing the frequency with which the model is recalibrated.

\begin{figure}[t]
\begin{center}
\includegraphics[width=.45\textwidth]{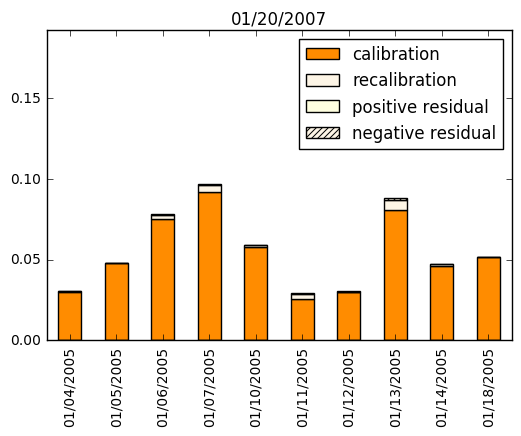}
\includegraphics[width=.45\textwidth]{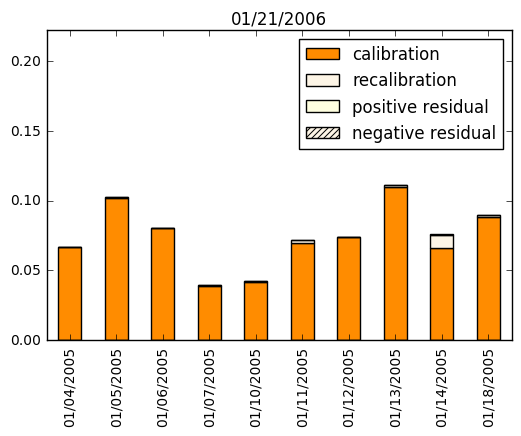}
\includegraphics[width=.45\textwidth]{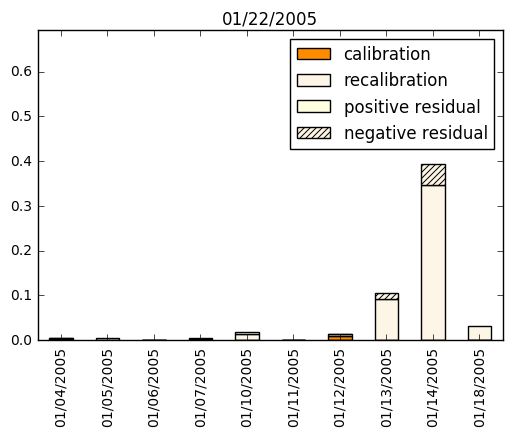}
\includegraphics[width=.45\textwidth]{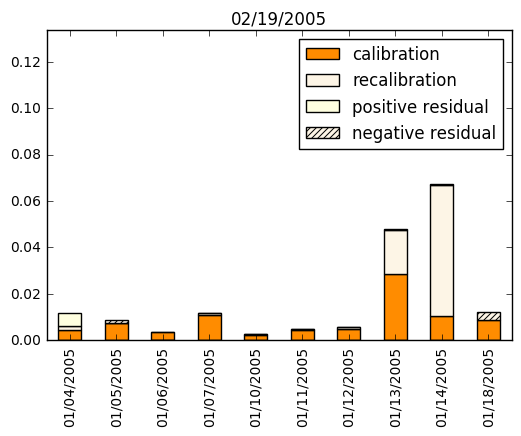}
\includegraphics[width=.45\textwidth]{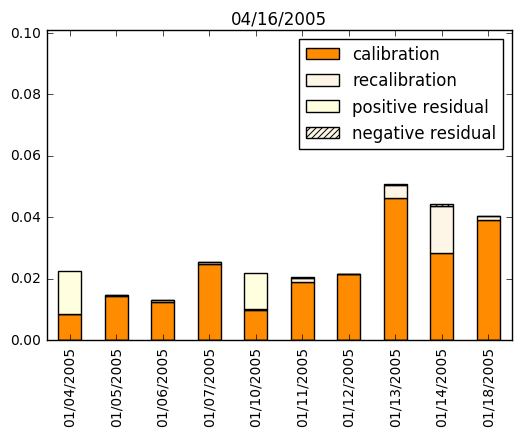}
\includegraphics[width=.45\textwidth]{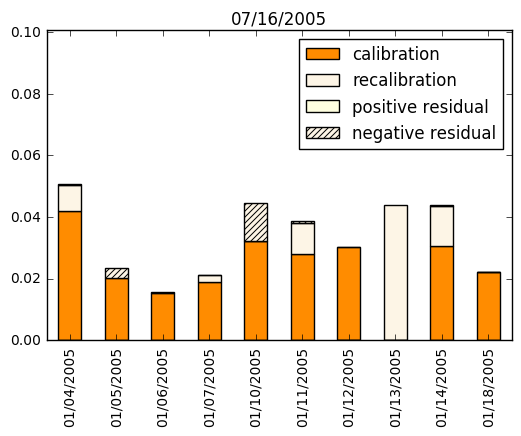}
\caption{Decomposition of model risks of AAPL options using the Black/Scholes model as the nominal model, for a selection of option maturity dates (given as the title of each graph).}\label{fig:AAPLdecomp}
\end{center}
\end{figure}
We start by evaluating the calibration, recalibration and aggregated model risks under a Black/Scholes model, i.e. where the underlying asset price is assumed to follow geometric Brownian motion, with dynamics under the risk--neutral measure given by
\begin{equation}\label{eq:BSSDE}
dS(t)=rS(t)dt+\sigma S(t)dW(t)
\end{equation}
where $r$ is the continuously compounded risk--free rate of interest and $\sigma$ is a constant volatility parameter. We note that in the Black/Scholes model we obtain a simple closed form expression for the recalibration risk defined in (\ref{eq:eta2}):
\begin{eqnarray} \label{eq:etaf}
	\eta_2=\left(\frac{\sigma^{*2}}{\sigma^2}-1\right)\left[\frac{1}{2}+\frac{T}{8}(\sigma^{*2}-\sigma^2)\right]+\ln\left(\frac{\sigma}{\sigma^*}\right)
\end{eqnarray}
where $\sigma^*$ is the correctly recalibrated Black/Scholes volatility parameter and $\sigma$ is the parameter value obtained in a previous calibration. This formula is a consequence of the log-normal distribution of returns assumed in the Black/Scholes model.

We can express the aggregate model risk as the sum of the calibration error, the recalibration risk and a residual. As noted in equation (\ref{eq:add}), the residual is zero if the likelihood ratios involved in the calibration and recalibration risks are two independent random variables. In practice, the residual usually takes a small non-zero value. In Figure \ref{fig:AAPLdecomp} we demonstrate the decomposition of the total model risk into the three components.\footnote{The vertical axis denotes the numerical value of the relative entropy.} Unsurprisingly (as it is well documented that the Black/Scholes model cannot fit the implied volatility ``smile'' observed in most options markets), we see that calibration error typically predominates.

In the Heston model, the dynamics (\ref{eq:BSSDE}) are extended to allow for stochastic volatility, i.e.
\begin{eqnarray}
dS(t) &=& rS(t)dt+\sigma S(t)dW_1(t)\\
d(\sigma^2(t)) &=& \kappa(\theta-\sigma^2(t))dt + \eta \sigma(t)dW_2(t)
\end{eqnarray}
This model involves two state variables, the underlying asset price $S(t)$ and the volatility $\sigma(t)$, and five model parameters: $r$, $\kappa$, $\theta$, $\eta$, $\rho$ where $\rho$ is the correlation coefficient between the two Wiener processes:
$$
d\bracket{W_1}{W_2}_t=\rho dt
$$
$r$ is the risk-free rate,\footnote{In our empirical application examples, we take the risk--free rate as one of the financial variables observed in the market, but we do not explicitly take into account interest rate risk in our empirical analysis. For the short--dated options considered here, interest rate risk is known to be of relatively little importance --- for a discussion of this issue, see e.g. \citeasnoun{CHENG2017} and the literature cited therein.} and $\kappa$, $\theta$ and $\eta$ relate to the volatility process, $\kappa$ being the rate of mean reversion, $\theta$ the long--run mean and $\eta$ the volatility of this process.

\begin{figure}[t]
\begin{center}
\includegraphics[width=.45\textwidth]{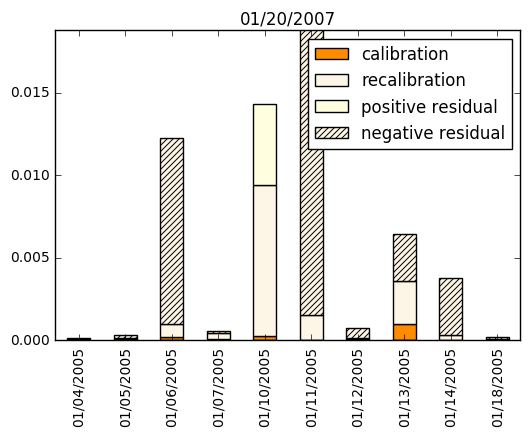}
\includegraphics[width=.45\textwidth]{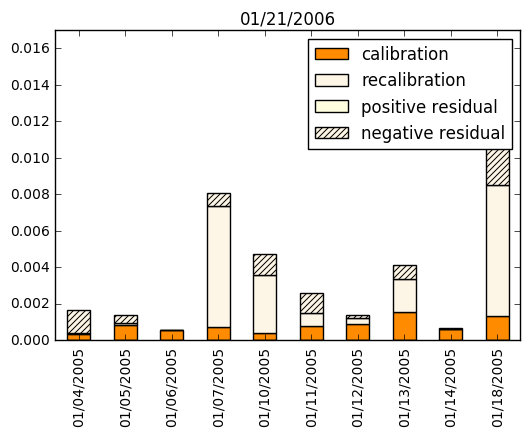}
\includegraphics[width=.45\textwidth]{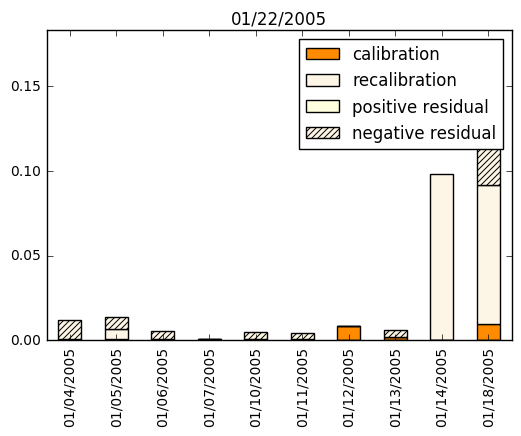}
\includegraphics[width=.45\textwidth]{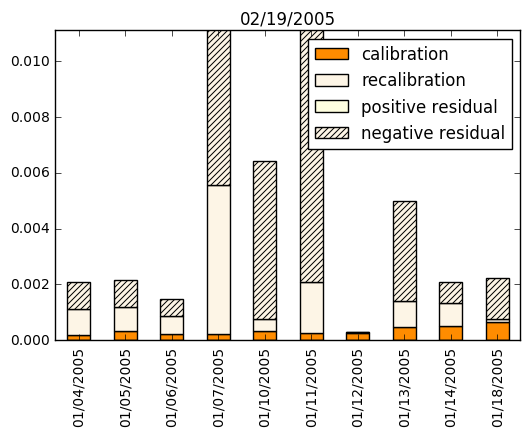}
\includegraphics[width=.45\textwidth]{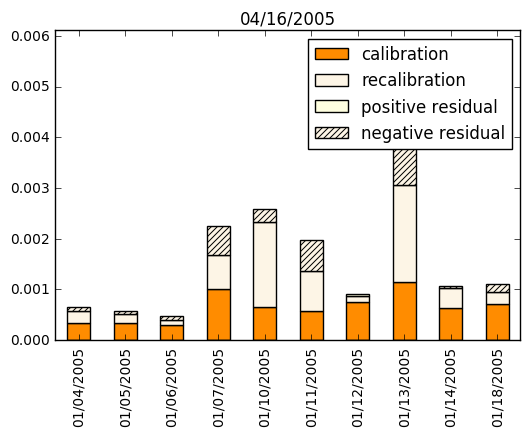}
\includegraphics[width=.45\textwidth]{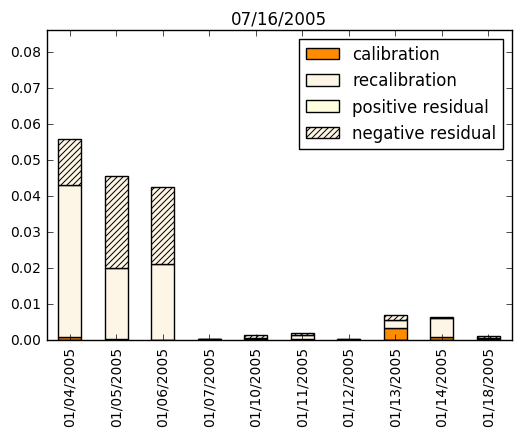}
\caption{Decomposition of model risks of GOOG options using the Heston model as the nominal model, for a selection of option maturity dates (given as the title of each graph).}\label{fig:GOOGdecomp}
\end{center}
\end{figure}
Following \citeasnoun{OZ:Gatheral:06}, the risk-neutral probability of exercise of a European call option with strike $K$ in the Heston model is given by
\begin{align}
P_0(x,v,\tau)=
\frac{1}{2}+\frac{1}{\pi}
\int_0^\infty \mathfrak{R}
\left[\frac{\exp\{C(u,\tau)\theta+D(u,\tau)v+iux\}}{iu}\right]du\,,
\end{align}
where $v$ is the current value of the volatility state variable $\sigma(t)$, $\tau=T-t$ is the time to maturity, and $x$ is the logarithmic forward moneyness
of the option, i.e.
\begin{eqnarray*}
x: &=& \ln\left(\frac{F_{t,T}}{K}\right)\\
F_{t,T} &=& \frac{S(t)}{B(t,T)}
\end{eqnarray*}
with $B(t,T)$ the time $t$ price of a zero bond maturing in $T$. Furthermore,
\begin{align}
C(u,\tau)=&\kappa\left[r_-\tau-\frac{2}{\eta^2}\ln\left(\frac{r_+-r_-e^{-d\tau}}{r_+-r_-}\right)\right]\\
D(u,\tau)=&r_-\frac{r_+-r_+e^{-d\tau}}{r_+-r_-e^{-d\tau}}\\
r_\pm = \frac{\beta\pm d}{\eta^2}&\qquad
d=\sqrt{\beta^2-2\alpha\eta^2}
\end{align}
Parameters $\alpha$ and $\beta$ are functions of $u$ (Fourier transform variable of $x$):
\begin{align}
\alpha(u)=& -\frac{u^2}{2}-i\frac{u}{2}\\
\beta(u)=& \kappa-\rho\eta iu
\end{align}
\begin{table}[t]
\begin{center}
{\scriptsize
	\begin{tabular}{cc|c|c|c|c|c|c|c|c|c|c}
		\multicolumn{12}{c}{Aggregate Model Risk}\\[5pt]
		\hline\hline
		\multicolumn{2}{c|}{\multirow{2}{*}{Risk measure}} &
		\multicolumn{5}{c|}{Black-Scholes} & \multicolumn{5}{|c}{Heston}\\
		\cline{3-12}
		& &
		1 day & 3 days & 1 week & 2 weeks & 1 quarter &
		1 day & 3 days & 1 week & 2 weeks & 1 quarter \\
		\hline
		\multicolumn{2}{c|}{mean} & 0.070& 0.071& 0.073& 0.073&0.085 & 0.037& 0.035& 0.038& 0.039& 0.046\\
		\multicolumn{2}{c|}{median} & 0.045& 0.047& 0.047& 0.051& 0.057& 0.005& 0.005& 0.005& 0.006&0.010\\
		\hline
		\multirow{4}{*}{Quantile} & 99\% & 0.474& 0.427& 0.471& 0.455&0.462 & 0.508& 0.462& 0.512&0.503&0.649\\
		 & 95\% & 0.212& 0.221& 0.221& 0.212& 0.251& 0.173& 0.169& 0.177&0.177&0.185\\
		 & 90\% & 0.158& 0.160& 0.165& 0.160& 0.192& 0.096& 0.097& 0.098&0.105&0.121\\
		 & 75\% & 0.092& 0.094& 0.095& 0.099& 0.112& 0.027& 0.027& 0.029&0.031&0.034\\
		\hline\\
		\multicolumn{12}{c}{Calibration Error}\\[5pt]
		\hline
		\multicolumn{2}{c|}{mean} & 0.055& 0.066& 0.069& 0.072& 0.084& 0.008& 0.026& 0.032& 0.036&0.045 \\
		\multicolumn{2}{c|}{median} & 0.038& 0.043& 0.045& 0.049& 0.057& 0.001& 0.003& 0.004& 0.005&0.009\\
		\hline
		\multirow{4}{*}{Quantile} & 99\% & 0.239& 0.397& 0.433& 0.443& 0.455& 0.163& 0.416& 0.496& 0.495&0.648\\
		 & 95\% &0.171& 0.207& 0.212& 0.210& 0.249& 0.015& 0.130& 0.158& 0.168&0.182\\
		 & 90\% & 0.135& 0.151& 0.158& 0.158& 0.191& 0.008& 0.065& 0.083&0.097&0.119\\
		 & 75\% & 0.075& 0.087& 0.091& 0.096& 0.111& 0.003& 0.012& 0.019&0.026&0.034\\
		\hline\\
		\multicolumn{12}{c}{Model Risk due to Recalibration}\\[5pt]
		\hline
		\multicolumn{2}{c|}{mean}& 0.024& 0.009& 0.005& 0.002& 0.001 & 0.057& 0.019& 0.011& 0.006& 0.001\\
		\multicolumn{2}{c|}{median} & 0.002& 0.001& 0.001& 0.000& 0.000& 0.004& 0.001& 0.001&0.001&0.000\\
		\hline
		\multirow{4}{*}{Quantile} & 99\% & 0.458& 0.196& 0.057& 0.030& 0.012& 0.630& 0.226& 0.117&0.067&0.011\\
		 & 95\% & 0.103& 0.034& 0.021& 0.010& 0.005& 0.315& 0.108& 0.062& 0.032&0.006\\
		 & 90\% & 0.056& 0.020& 0.011& 0.006& 0.002& 0.164& 0.052& 0.031&0.016&0.004\\
		 & 75\% & 0.013& 0.006& 0.004& 0.002& 0.001& 0.055& 0.016& 0.011&0.005&0.001\\
		\hline\hline
	\end{tabular}}\vspace*{1.5ex}

\end{center}
\caption{Model risks (in terms of relative entropy) under different models and recalibration frequencies for AAPL.}\label{AAPLtbl}
\end{table}
It is noted that $P_0=E^\mathbb{Q}(\mathbbm{1}_{S_T>K})$ since by definition $P_0$ is the probability of exercise. The probability density function of the risk-neutral measure is therefore obtained:
\begin{align}
p(S_T)=&-\left.\frac{\partial P_0}{\partial K}\right|_{K=S_T}=-\frac{\partial P_0}{\partial x}\left.\frac{\partial x}{\partial K}\right|_{K=S_T}\\
=&\frac{1}{2\pi S_T}\int_{-\infty}^\infty
\exp\left[C(u,\tau)\theta+D(u,\tau)v+iu\ln\left(\frac{S_t}{B(t,T)S_T}\right)\right]du
\end{align}
For simplicity $e^y$ denotes the ratio of the forward price at $t$ to its spot price at maturity $T$, i.e. $e^y=B(t,T)S_T/S_t$, we derive the risk-neutral probability with respect to $y$:
\begin{align}
p(y)=&p(S_T)\left|\frac{\partial y}{\partial S_T}\right|^{-1}\\
=&\frac{1}{2\pi }\int_{-\infty}^\infty
\exp\left[C(u,\tau)\theta+D(u,\tau)v-iuy\right]du
\end{align}
$p(y)$ can be calculated by fast Fourier transform (FFT).

The decomposition of the total model risk into the three components (the components due calibration and recalibration, and the positive or negative residual measuring the departure from independence between the first two components) when using the Heston model as the baseline is given in Figure \ref{fig:GOOGdecomp}. Again, since it is well documented that the Heston model can fit observed option prices better than Black/Scholes, it is unsurprising that in this case the relative entropy measuring calibration error is much lower --- however, already in this set of example days it is evident that this comes at a price of increased model risk due to recalibration.

\begin{table}[t]
\begin{center}
{\scriptsize
	\begin{tabular}{cc|c|c|c|c|c|c|c|c|c|c}
		\multicolumn{12}{c}{Aggregate Model Risk}\\[5pt]
		\hline\hline
		\multicolumn{2}{c|}{\multirow{2}{*}{Risk measure}} &
		\multicolumn{5}{c|}{Black-Scholes} & \multicolumn{5}{|c}{Heston}\\
		\cline{3-12}
		& &
		1 day & 3 days & 1 week & 2 weeks & 1 quarter &
		1 day & 3 days & 1 week & 2 weeks & 1 quarter \\
		\hline
		\multicolumn{2}{c|}{mean} & 0.165& 0.168& 0.165& 0.165& 0.188& 0.115& 0.105& 0.113& 0.114&0.127 \\
		\multicolumn{2}{c|}{median} & 0.109& 0.111& 0.112& 0.115& 0.138& 0.053& 0.050& 0.055& 0.053&0.059\\
		\hline
		\multirow{4}{*}{Quantile} & 99\% & 0.705& 0.722& 0.722& 0.699& 0.728& 0.726& 0.668& 0.711&0.699&0.744\\
		 & 95\% & 0.519& 0.530& 0.521& 0.496& 0.549& 0.475& 0.438& 0.455&0.467&0.560\\
		 & 90\% & 0.393& 0.391& 0.387& 0.385& 0.408& 0.330& 0.287&0.320&0.324&0.380\\
		 & 75\% & 0.219& 0.223& 0.214& 0.215& 0.256& 0.145& 0.137& 0.142&0.153&0.157\\
		\hline\\
		\multicolumn{12}{c}{Calibration Error}\\[5pt]
		\hline
		\multicolumn{2}{c|}{mean} & 0.125& 0.155& 0.158& 0.162& 0.187& 0.055& 0.083& 0.103& 0.107&0.126\\
		\multicolumn{2}{c|}{median} & 0.082& 0.102& 0.106& 0.111& 0.137& 0.006& 0.022& 0.043& 0.046&0.058\\
		\hline
		\multirow{4}{*}{Quantile} & 99\% & 0.655& 0.701& 0.709& 0.696& 0.728& 0.626& 0.651& 0.706&0.688&0.744\\
		 & 95\% &0.400& 0.482& 0.507& 0.494& 0.548& 0.329& 0.381& 0.438&0.445&0.557\\
		 & 90\% & 0.295& 0.359& 0.362& 0.379& 0.408& 0.172& 0.250& 0.298&0.306&0.375\\
		 & 75\% & 0.171& 0.208& 0.207& 0.212& 0.255& 0.034& 0.103& 0.127&0.141&0.155\\
		\hline\\
		\multicolumn{12}{c}{Model Risk due to Recalibration}\\[5pt]
		\hline
		\multicolumn{2}{c|}{mean}& 0.036& 0.014& 0.008& 0.004& 0.001 & 0.119& 0.044& 0.012& 0.012&0.003 \\
		\multicolumn{2}{c|}{median} & 0.004& 0.002& 0.001& 0.001& 0.000& 0.043& 0.016& 0.003&0.003&0.001\\
		\hline
		\multirow{4}{*}{Quantile} & 99\% & 0.405& 0.172& 0.091& 0.050& 0.010& 0.757& 0.255& 0.087&0.077&0.012\\
		 & 95\% & 0.173& 0.066& 0.035& 0.018& 0.004& 0.543& 0.187& 0.056&0.064&0.011\\
		 & 90\% & 0.109& 0.038& 0.022& 0.012& 0.003& 0.411& 0.151& 0.034&0.045&0.009\\
		 & 75\% & 0.036& 0.013& 0.008& 0.004& 0.001& 0.142& 0.052&0.016&0.013&0.003\\
		\hline\hline
	\end{tabular}}\vspace*{1.5ex}

\end{center}
\caption{Model risks (in terms of relative entropy) under different models and recalibration frequencies for GOOG.}\label{GOOGtbl}
\end{table}
These observations are reinforced when we consider aggregate model risk, calibration error and model risk due to recalibration over the entire sample period, as presented in Tables \ref{AAPLtbl} and \ref{GOOGtbl}. Note that the absolute numbers refer to relative entropy and thus lack direct financial interpretation --- what matters are the relative values when comparing the model across models and different recalibration frequencies, in particular when considering the aggregate model risk. Here, we consider recalibrating the Black/Scholes and Heston models either daily, every three days, every week, every two weeks, or every quarter year. We see that recalibrating more frequently has little effect on the aggregate model risk, neither when using the Black/Scholes model nor when using the Heston model. Essentially, recalibrating more frequently simply shifts calibration error into model risk due to recalibration,\footnote{Note that on days on which we do not recalibrate, the model risk due to recalibration is zero, because (consistent with the model assumptions) we are keeping previously calibrated parameters unchanged --- so on those days aggregate model risk is entirely due to calibration error (which increases because the fit to market prices deteriorates when we do not recalibrate).} highlighting the dangers in the common practice of focusing solely on the calibration of derivative pricing models, at the expense of all other sources of model risk.

In addition, we observe that if we are interested in ``robustness'' at a high level of confidence (looking at, say, the 99\% quantile of aggregate model risk), moving from Black/Scholes to Heston also does not appear to deliver any advantage (it does yield some improvement at lower quantiles, or average or median, aggregate model risk). This means that when high levels of confidence are required, any gain in calibration accuracy delivered by the Heston model is offset by higher model risk due to recalibration. One should note that this last point holds even \emph{before} considering Type 3 model risk, which may well be worse when additional state variables are introduced (as in the Heston model). For these results, in Tables \ref{AAPLtbl} and \ref{GOOGtbl} we calculated means, medians and quantiles across all available option maturities. If we consider only particular maturity ``buckets'', the same qualitative conclusions are evident --- Tables \ref{AAPLtbltwo} and \ref{GOOGtbltwo} illustrate this in the case of daily recalibration.

\begin{table}[t]
\begin{center}
{\scriptsize
	\begin{tabular}{cc|c|c|c|c|c|c|c|c}
		\multicolumn{10}{c}{Aggregate Model Risk}\\[5pt]
		\hline\hline
		\multicolumn{2}{c|}{\multirow{2}{*}{Risk measure}} &
		\multicolumn{4}{c|}{Black-Scholes} & \multicolumn{4}{|c}{Heston}\\
		\cline{3-10}
		& &
		all & 0-0.2 year & 0.2-0.7 year & $>$0.7 year &
		all & 0-0.2 year & 0.2-0.7 year & $>$0.7 year \\
		\hline
		\multicolumn{2}{c|}{mean} & 0.070&0.041 &0.066 &0.109 &0.037 &0.039 &0.024 &0.046  \\
		\multicolumn{2}{c|}{median} & 0.045& 0.021& 0.052& 0.081& 0.005& 0.004& 0.004&0.008\\
		\hline
		\multirow{4}{*}{Quantile} & 99\% &0.474 &0.471 &0.266 &0.567 &0.508 &0.590 &0.302&0.472\\
		 & 95\% &0.212 &0.120 &0.185 &0.282 &0.173 &0.243 &0.091&0.191\\
		 & 90\% &0.158 &0.081 &0.143 &0.213 &0.096 &0.074 &0.066&0.142\\
		 & 75\% & 0.092&0.047 &0.090 &0.141 &0.027 &0.021 &0.015&0.054\\
		\hline\\
		\multicolumn{10}{c}{Calibration Error}\\[5pt]
		\hline
		\multicolumn{2}{c|}{mean} & 0.055& 0.026& 0.057& 0.087& 0.008& 0.012& 0.004& 0.007\\
		\multicolumn{2}{c|}{median} & 0.038& 0.014& 0.048& 0.068& 0.001& 0.001& 0.001&0.001\\
		\hline
		\multirow{4}{*}{Quantile} & 99\% & 0.239& 0.157& 0.230& 0.289& 0.163& 0.072&0.052&0.116\\
		 & 95\% & 0.171& 0.091& 0.157& 0.212& 0.015& 0.021&0.011&0.017\\
		 & 90\% & 0.135& 0.062& 0.126& 0.181& 0.008& 0.006&0.007&0.010\\
		 & 75\% & 0.075& 0.036& 0.077& 0.128& 0.003& 0.002&0.003&0.003\\
		\hline\\
		\multicolumn{10}{c}{Model Risk due to Recalibration}\\[5pt]
		\hline
		\multicolumn{2}{c|}{mean} & 0.024& 0.020& 0.013& 0.038& 0.057& 0.062& 0.048&0.060 \\
		\multicolumn{2}{c|}{median} & 0.002& 0.002& 0.001& 0.003& 0.004& 0.004& 0.003&0.009\\
		\hline
		\multirow{4}{*}{Quantile} & 99\% & 0.458& 0.506& 0.116& 0.662& 0.630& 0.702&0.536&0.512\\
		 & 95\% &0.103& 0.066& 0.066& 0.169& 0.315& 0.443& 0.274&0.247\\
		 & 90\% & 0.056& 0.030& 0.042& 0.102& 0.164& 0.223&0.127&0.160\\
		 & 75\% & 0.013& 0.009& 0.010& 0.031& 0.055& 0.030&0.048&0.091\\
		\hline\hline
	\end{tabular}}\vspace*{1.5ex}

\end{center}
\caption{Model risks (in terms of relative entropy) under different models, by maturity buckets, for AAPL (daily recalibration frequency).}\label{AAPLtbltwo}
\end{table}

\begin{table}[t]
\begin{center}
{\scriptsize
	\begin{tabular}{cc|c|c|c|c|c|c|c|c}
		\multicolumn{10}{c}{Aggregate Model Risk}\\[5pt]
		\hline\hline
		\multicolumn{2}{c|}{\multirow{2}{*}{Risk measure}} &
		\multicolumn{4}{c|}{Black-Scholes} & \multicolumn{4}{|c}{Heston}\\
		\cline{3-10}
		& &
		all & 0-0.2 year & 0.2-0.7 year & $>$0.7 year &
		all & 0-0.2 year & 0.2-0.7 year & $>$0.7 year \\
		\hline
		\multicolumn{2}{c|}{mean} & 0.165& 0.167& 0.162& 0.164& 0.115& 0.115& 0.120&0.109\\
		\multicolumn{2}{c|}{median} & 0.109& 0.104& 0.099& 0.120& 0.053& 0.052& 0.057&0.048\\
		\hline
		\multirow{4}{*}{Quantile} & 99\% & 0.705& 0.736& 0.709& 0.661& 0.726& 0.714&0.707&0.736\\
		 & 95\% &0.519& 0.527& 0.551& 0.459& 0.475& 0.468& 0.500&0.476\\
		 & 90\% & 0.393& 0.417& 0.382& 0.371& 0.330& 0.351&0.322&0.304\\
		 & 75\% & 0.219& 0.221& 0.216& 0.218& 0.145& 0.149&0.155&0.136\\
		\hline\\
		
		\multicolumn{10}{c}{Calibration Error}\\[5pt]
		\hline
		\multicolumn{2}{c|}{mean} & 0.125& 0.121& 0.123& 0.132& 0.055& 0.055& 0.052& 0.057\\
		\multicolumn{2}{c|}{median} & 0.082& 0.074& 0.075& 0.099& 0.006& 0.005& 0.006&0.006\\
		\hline
		\multirow{4}{*}{Quantile} & 99\% & 0.655& 0.662& 0.650& 0.639& 0.626& 0.658&0.531&0.665\\
		 & 95\% & 0.400& 0.403& 0.423& 0.391& 0.329& 0.349&0.311&0.330\\
		 & 90\% & 0.295& 0.298& 0.301& 0.285& 0.172& 0.155&0.185&0.171\\
		 & 75\% & 0.171& 0.162& 0.160& 0.176& 0.034& 0.031&0.034&0.034\\
		\hline\\
		
		\multicolumn{10}{c}{Model Risk due to Recalibration}\\[5pt]
		\hline
		\multicolumn{2}{c|}{mean} & 0.036&0.038&0.039&0.033&0.119&0.117&0.117&0.124\\
		\multicolumn{2}{c|}{median} & 0.004& 0.005& 0.003& 0.005& 0.043& 0.043& 0.037&0.049\\
		\hline
		\multirow{4}{*}{Quantile} & 99\% & 0.405&0.419&0.479&0.284&0.757&0.746&0.767&0.754\\
		 & 95\% & 0.173& 0.176& 0.188& 0.155& 0.543& 0.510& 0.558&0.554\\
		 & 90\% &0.109 &0.111&0.115&0.098&0.411 &0.390&0.418&0.418\\
		 & 75\% & 0.036&0.038&0.032&0.036&0.142&0.144&0.137&0.146\\
		\hline\hline
	\end{tabular}}\vspace*{1.5ex}

\end{center}
\caption{Model risks (in terms of relative entropy) under different models, by maturity buckets, for GOOG (daily recalibration frequency).}\label{GOOGtbltwo}
\end{table}

\section{Conclusion}\label{conclusion}
Under our approach, less relative entropy implies less model risk, and we are able to evaluate two hitherto disparate sources of model risk (calibration error and model risk due to recalibration) in a unified fashion, and examine the potential trade--off between the two. We have considered a simple choice between two models, and between different recalibration frequencies. ``Putting a number on model risk'' by calculating quantiles for the maximum model risk (quantified by relative entropy) over a time series of market data allows one to assess the added value (if any) of more complicated stochastic models of financial markets.

In our application, we are deliberately prioritising the minimisation of calibration error, as this is congruent to the (often exclusive) focus of practitioners on calibration error (with little or no regard to model risk due to recalibration).\footnote{If desired, one could reformulate this approach to prioritise the minimisation of aggregate model risk, or of model risk due to recalibration.} Even in this case, we see that by including recalibration as one of the sources of aggregate model risk, recalibrating a model frequently to a changing market simply interchanges one source of model risk for another, and more complicated stochastic models may well underperform when aggregate model risk is taken into account.

\bibliography{modelrisk}

\end{document}